\documentclass[fleqn,usenatbib]{mnras}

\usepackage{newtxtext,newtxmath}
\usepackage[T1]{fontenc}

\DeclareRobustCommand{\VAN}[3]{#2}
\let\VANthebibliography\thebibliography
\def\thebibliography{\DeclareRobustCommand{\VAN}[3]{##3}\VANthebibliography}

\usepackage{graphicx}	
\usepackage{amsmath}	
\usepackage{amssymb}	
\usepackage{multirow}
\usepackage{caption}
\usepackage{todonotes}

\newcounter{savefootnote}
\newcounter{symfootnote}
\newcommand{\symfootnote}[1]{%
   \setcounter{savefootnote}{\value{footnote}}%
   \setcounter{footnote}{\value{symfootnote}}%
  \ifnum\value{footnote}>8\setcounter{footnote}{0}\fi%
   \let\oldthefootnote=\thefootnote%
   \renewcommand{\thefootnote}{\fnsymbol{footnote}}%
   \footnote{#1}%
   \let\thefootnote=\oldthefootnote%
   \setcounter{symfootnote}{\value{footnote}}%
   \setcounter{footnote}{\value{savefootnote}}%
}

\title{A deep learning view of the census of galaxy clusters in IllustrisTNG}
\author[Y.\ Su et al.]
{Y.\ Su,$^{1}$\thanks{E-mail:ysu262@uky.edu}
Y.\ Zhang,$^{1,2}$
G.\ Liang,$^{1,2}$
J.\ A.\ ZuHone,$^{3}$
D.\ J.\ Barnes,$^{4}$
N.\ B.\ Jacobs,$^{2}$
\newauthor
M.\ Ntampaka,$^{3,5}$
W.\ R.\ Forman,$^{3}$
P.\ E.\ J.\ Nulsen,$^{3}$
R.\ P.\ Kraft,$^{3}$
and C.\ Jones.$^{3}$
\\
$^{1}$Department of Physics and Astronomy, University of Kentucky, 505 Rose Street, Lexington, KY 40506, USA\\
$^{2}$Department of Computer Science, University of Kentucky, 329 Rose Street, Lexington, KY 40506, USA\\
$^{3}$Center for Astrophysics $|$ Harvard \& Smithsonian, Cambridge, MA 02138, USA\\
$^{4}$Department of Physics, Kavli Institute for Astrophysics and Space Research, Massachusetts Institute of Technology, Cambridge, MA 02139, USA\\
$^{5}$Harvard Data Science Initiative, Harvard University, Cambridge, MA 02138, USA
}
\date{Accepted XXX. Received YYY; in original form ZZZ}

\pubyear{2015}

\begin{document}
\label{firstpage}
\pagerange{\pageref{firstpage}--\pageref{lastpage}}
\maketitle

\begin{abstract}

The origin of the diverse population of galaxy clusters remains an unexplained aspect of large-scale structure formation and cluster evolution.   
We present a novel method of using X-ray images to identify cool core (CC), weak cool core (WCC), and non cool core (NCC) clusters of galaxies, that are defined by their central cooling times. 
We employ a convolutional neural network, ResNet-18, which is commonly used for image analysis, to classify clusters.
We produce mock {\sl Chandra} X-ray observations for a sample of 318 massive clusters drawn from the {\it IllustrisTNG} simulations.
The network is trained and tested with low resolution mock {\sl Chandra} images covering 
a central 1 Mpc square for the clusters in our sample. 
Without any spectral information, the deep learning algorithm is able to identify CC, WCC, and NCC clusters, achieving balanced accuracies (BAcc) 
of 92\%, 81\%, and 83\%, respectively. 
The performance is superior to classification by conventional methods using central gas densities, with an average ${\rm BAcc}=81\%$, or surface brightness concentrations, giving ${\rm BAcc}=73\%$.
We use Class Activation Mapping to localize discriminative regions for the classification decision. From this analysis, we observe that the network has utilized regions from cluster centers out to $r\approx300$\,kpc and $r\approx500$\,kpc to identify CC and NCC clusters, respectively. 
It may have recognized features in the intracluster medium that are 
associated with AGN feedback and disruptive major mergers.

\end{abstract}

\begin{keywords}
X-rays: galaxies: clusters -- galaxies: clusters: intracluster medium -- methods: data analysis
\end{keywords}



\section{Introduction}
As the product of hierarchical structure formation,
clusters of galaxies are the largest gravitationally collapsed objects in the Universe, carrying valuable information on the nature of dark matter and dark energy.
Clusters of galaxies contain vast reservoirs of intracluster medium (ICM), radiating vigorously in X-rays,
providing unique laboratories to study the cooling and heating of the hot baryons  
and the astrophysical processes that shape their thermodynamical properties.

Galaxy clusters are conventionally divided into three categories: cool core (CC), weak cool core (WCC), and non cool core (NCC) based on their core properties. CC clusters feature a sharp X-ray emission peak associated with a dense, cool, and enriched core \citep{sanders04}. The gas cooling time at centres of CC clusters is much shorter than the Hubble time. 
High sensitivity X-ray observations provided by {\sl Chandra} and {\sl XMM-Newton} reveal 
interactions between the
active galactic nuclei (AGN) at the centres of the brightest cluster galaxies (BCG) and the ambient ICM manifested by X-ray cavities, jets, and shocks 
\citep[e.g.,][]{2012ARA&A..50..455F, 2015ApJ...805..112R,2017ApJ...847...94S}, which could pump additional energy into the ICM and compensate for the radiative losses.   
In contrast, the gaseous, thermal, and chemical distributions of NCC clusters are relatively homogeneous over the inner region of a cluster. WCC clusters, often featuring a remnant cool core, appear to be an intermediate class (and possibly a transitional phase) between CC and NCC clusters~\citep{Su16, Markevitch03}.    

The origin of different populations of galaxy clusters has been a subject of debate for decades. In the prevailing model, a cool core is considered to be the natural state resulting from radiative cooling. Major mergers may have disrupted cluster cool cores and created NCC clusters, while CC clusters have only experienced minor or off-axis mergers. 
This interpretation is supported by X-ray observations showing that CC clusters appear to have a more symmetric morphology than NCC clusters~\citep{Buot96, Lovisari17}. Radio observations also reveal that clusters that host large scale diffuse synchrotron emissions,
suggesting that they have undergone a recent merger,
are predominantly NCC clusters~\citep{2011A&A...532A.123R}. However, CC and NCC clusters do not appear to have different gas properties at large radii~\citep{2020arXiv200701084G, 2019A&A...627A..19G}.
Conflicting results have also emerged in numerical simulations as to whether mergers are capable of transforming CC clusters into NCC clusters~\citep{Poole08, Barnes18, Rasia15}.  
In an alternative scenario, 
the presence (or absence) of a cool core is determined by the physical conditions and mechanisms at cluster centres,
e.g., the level of thermal conduction~\citep{Cavagnolo08, Voit08} and precipitation~\citep{2015Natur.519..203V}, 
the power of AGN outburts~\citep{2010ApJ...717..937G},
or the combined effect of mergers and AGN activity~\citep{2020arXiv200106532C}. X-ray observations indicate that gas
properties of cluster cores display little evolution over the last 10 Gyr, suggesting that
thermal equilibrium and feedback processes in cluster cores have been in place since the early Universe~\citep{Ghirardini20, 2017ApJ...843...28M, 
2015ApJ...805...35H, 2019ApJ...881...98S}.

It is desirable to obtain a complete and unbiased picture of galaxy clusters to understand the origin of their diversity, the interplay between the ICM and AGN feedback, and the formation and evolution of large scale structure.  
Flux-limited X-ray-selected samples are biased towards CC clusters as their centres are X-ray brighter than NCC clusters at a given cluster mass~\citep{Eckert11, Hudson10}. Recent Sunyaev--Zel'dovich (SZ) surveys provide nearly unbiased mass-limited samples of galaxy clusters. It was found that two-thirds of the Planck clusters are NCC or WCC~\citep{Felipe17, Rossetti17}. 

Ongoing and future extragalactic surveys such as eROSITA, SPT-3G, and LSST are designed to detect $\sim100,000$ clusters, allowing the model-independent determination of cosmological parameters~\citep{Haiman01}.
Modern data analysis techniques can be utilized to efficiently characterize the cluster properties across the electromagnetic spectrum. Machine learning tools have been applied to reduce errors in galaxy cluster X-ray masses~\citep{2019ApJ...884...33G, Ntampaka19}, dynamical masses~\citep{2015ApJ...803...50N, 2016ApJ...831..135N,  2019ApJ...887...25H, 2020arXiv200305951K}, SZ masses~\citep{2020arXiv200306135G}, lensing analyses \citep{2020arXiv200513985G, 2020MNRAS.491.5301S}, and to model micro-calorimeter X-ray spectra~\citep{2018MNRAS.475.4739I}.  
These techniques offer flexibility to take advantage of complicated correlations, well suited for mining large datasets and extracting information in the observational data that is inaccessible by conventional methods.

We present a deep learning approach to characterizing the thermodynamic structures of clusters of galaxies. The paper is structured as follows. In Section~\ref{sec:method},
we describe the IllustrisTNG simulations, the mock {\sl Chandra} observations, and the network architecture. We present the predicted cluster type classifications in Section~\ref{sec:results}. We discuss the implication of this work in Section~\ref{sec:discussion}, and conclude in Section~\ref{sec:conclusion}.

\section{Methods}
\label{sec:method}

\subsection{IllustrisTNG clusters}
\label{sec:tng}
The IllustrisTNG project includes a series of state-of-the-art cosmological magnetohydrodynamical simulations of galaxy formation \citep{nelson2018illustristng, Nelson_2017, Naiman_2018, Marinacci_2018} . 
It is a successor to the original Illustris simulation \citep{Vogelsberger14}. 
IllustrisTNG utilizes both large volumes and high resolutions, which reproduces relations between black hole masses and the properties of their host galaxies \citep{Li19}, the metal abundance of the ICM \citep{Vogelsberger17}, and the cosmic large scale structures \citep{Springel17}. 
TNG300 is the largest simulation volume in IllustrisTNG, containing a simulated cubic volume of (300\,Mpc)$^3$ with a baryonic mass resolution of $7.6\times10^6$\,$M_{\odot}$~\citep{2019MNRAS.490.3234N}, providing a rich and diverse collection of collapsed halos \citep{Pillepich17}.
The simulations use a cosmological model based on the constraints of \citet{Planck16} with $\Omega_{\rm m}=0.3089$, $\Omega_{\Lambda}=0.6911$, and $H_0=67.74$\,km\,s$^{-1}$\,Mpc$^{-1}$.

\begin{figure}
	\includegraphics[width=0.49\textwidth]{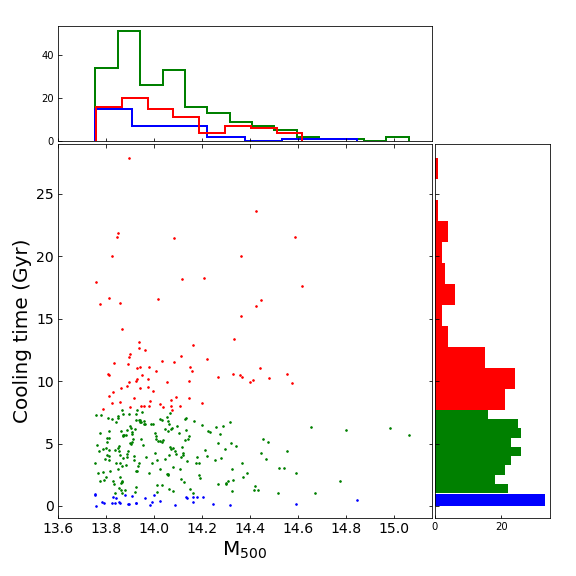}
    \caption{Distributions of central cooling times and log$M_{500}/M_{\odot}$ of TNG300 clusters in our sample. Their central cooling times are in the range of $0.012-27.85$\,Gyr. We define CC and NCC clusters as those with cooling times shorter than 1\,Gyr and longer than 7.7\,Gyr, respectively. Clusters with $1<t_{\rm cool}<7.7$\,Gyr are defined as WCC clusters. Clusters in our sample have $M_{500}$ in the range of $10^{13.75-15.06}M_{\odot}$.}
    \label{fig:hist}
\end{figure}

We select galaxy clusters with a total mass within $R_{500}$\symfootnote{$R_{\Delta}$ is the radius within which the overdensity of
the galaxy cluster is ${\Delta}$ times the critical density of the Universe.} above  
$M_{500} = 10^{13.75}$\,M$_{\odot}$ using the Friends-of-Friends algorithm ~\citep{1985ApJ...292..371D} from the $z=0$ snapshot in the TNG300 simulation, which forms an unbiased mass-limited sample of 318 massive clusters. 
A detailed analysis of the cluster populations in TNG300 is presented in \citet{Barnes18}. 
The radiative cooling time is defined as
\begin{equation}
t_{\rm cool}=\frac{3}{2}\frac{(n_e+n_i)k_BT}{n_en_i\Lambda(T,Z)}
\label{eq:tc}
\end{equation}
where $n_e$ and $n_i$ are the number densities of electrons and ions, respectively; $k_B$ is Boltzmann constant and $T$ is the gas temperature; $\Lambda$, the cooling
function, is determined by the plasma temperature and metallicity. 
Following \citet{Barnes18}, we calculate the average $t_{\rm cool}$ from a 3D volume within $0.012\,R_{500}$.
CC clusters are defined as those with $t_{\rm cool}<1$\,Gyr, an observation-based threshold for the presence of multi-phase gas likely due to the thermally unstable cooling. NCC clusters are those with $t_{\rm cool}>7.7$\,Gyr, corresponding to a lookback time to $z\approx1$ and representing the period since the last major merger. 
Clusters with $t_{\rm cool}$ between 1\,Gyr and 7.7\,Gyr are classified as WCC clusters. 
Such divisions for CC, WCC, and NCC clusters are commonly adopted in practice~\citep[e.g.,][]{McDonald13, Hudson10, 2017ApJ...851...66H, Barnes18}.
10\%, 61\%, 29\% of clusters in our sample are CC, WCC, and NCC, respectively. 
Distributions of the masses and cooling times of clusters in our sample are shown in Figure~\ref{fig:hist}.

\begin{figure*}
	\includegraphics[width=0.8\textwidth]{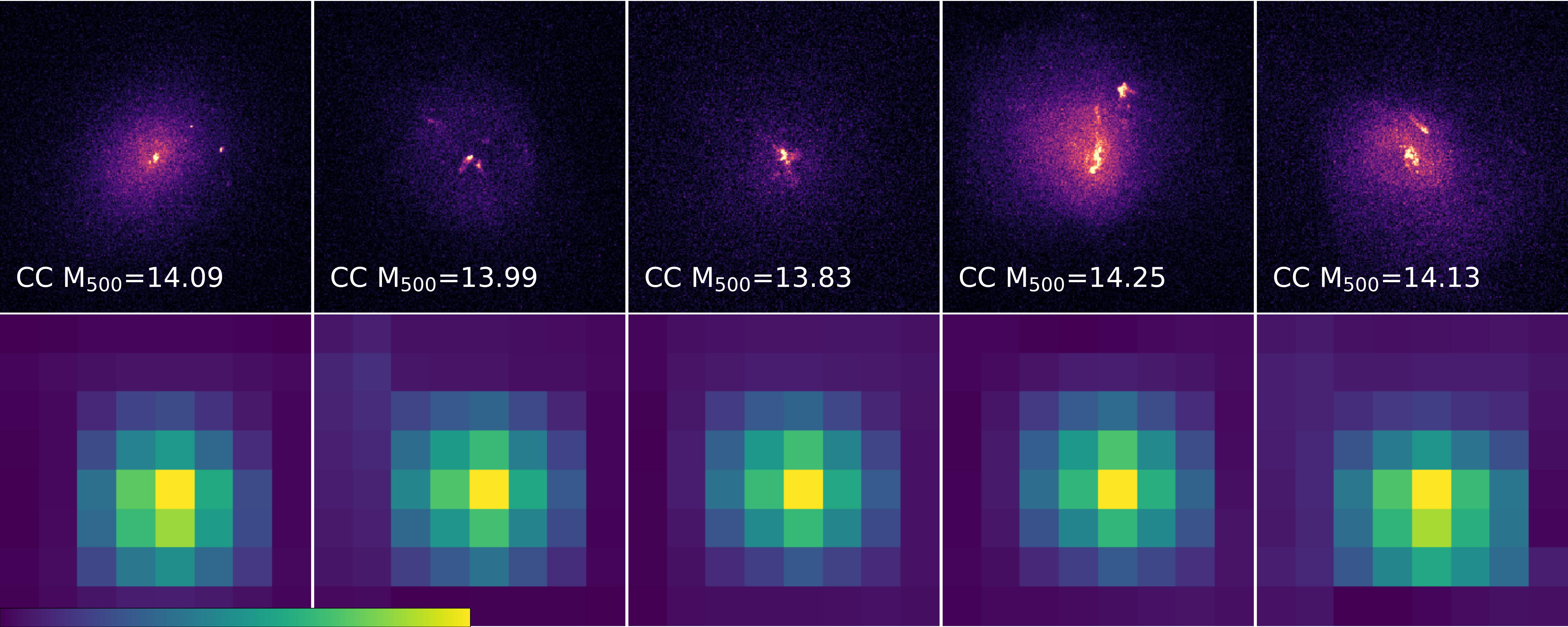}	
    \caption{{\bf top}: Example images of mock {\sl Chandra} observations of cool core clusters. Each image covers a D=1\,Mpc square region. {\bf bottom}: class activation maps highlight the discriminative regions in an image for the CNN to classify that image into a category. Each map corresponds to the above input image. All these clusters are predicted correctly with a probability above 0.9. The network has utilized radial ranges more extended than $r<0.012$\,$R_{500}$ (where the central cooling time and density are measured) to identify CC clusters.}
    \label{fig:cc_cam}
\end{figure*}

\begin{figure*}
	\includegraphics[width=0.8\textwidth]{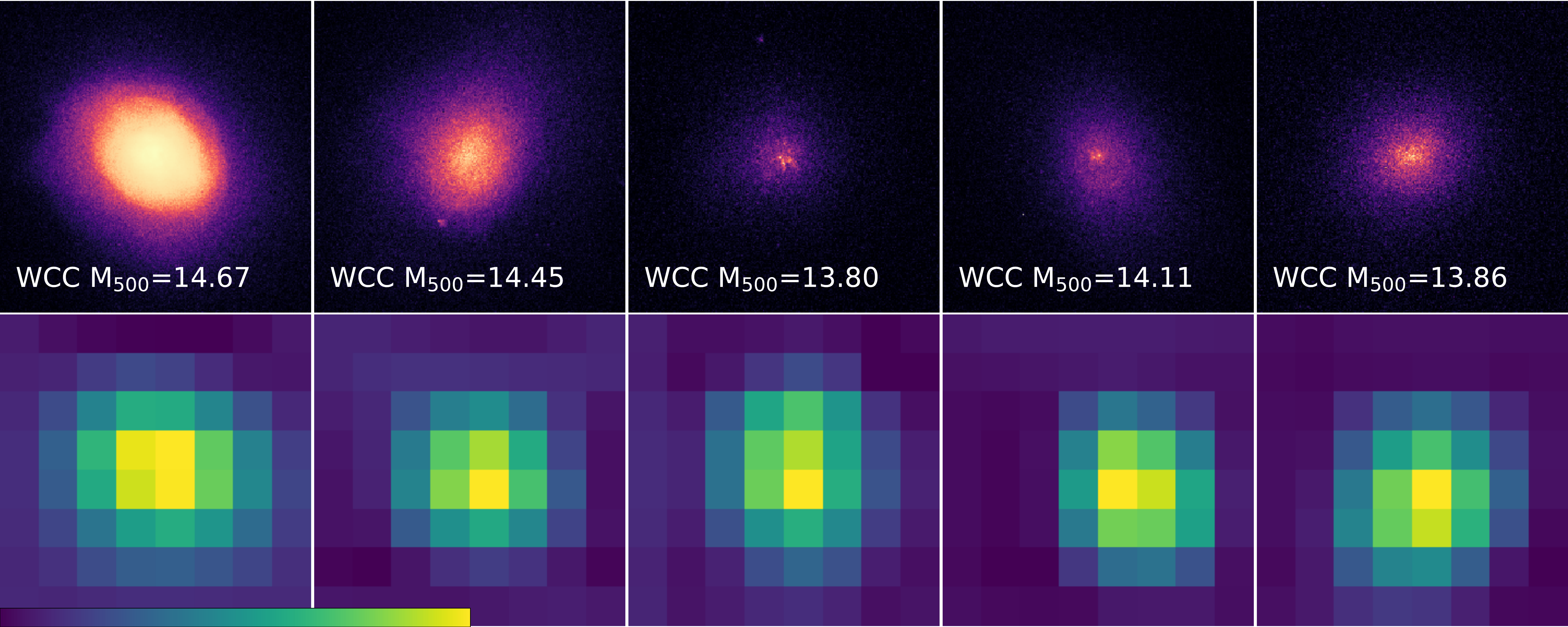}	
    \caption{Same as Figure~\ref{fig:cc_cam} but for weak cool core clusters.}
    \label{fig:wcc_cam}
\end{figure*}

\begin{figure*}
	\includegraphics[width=0.8\textwidth]{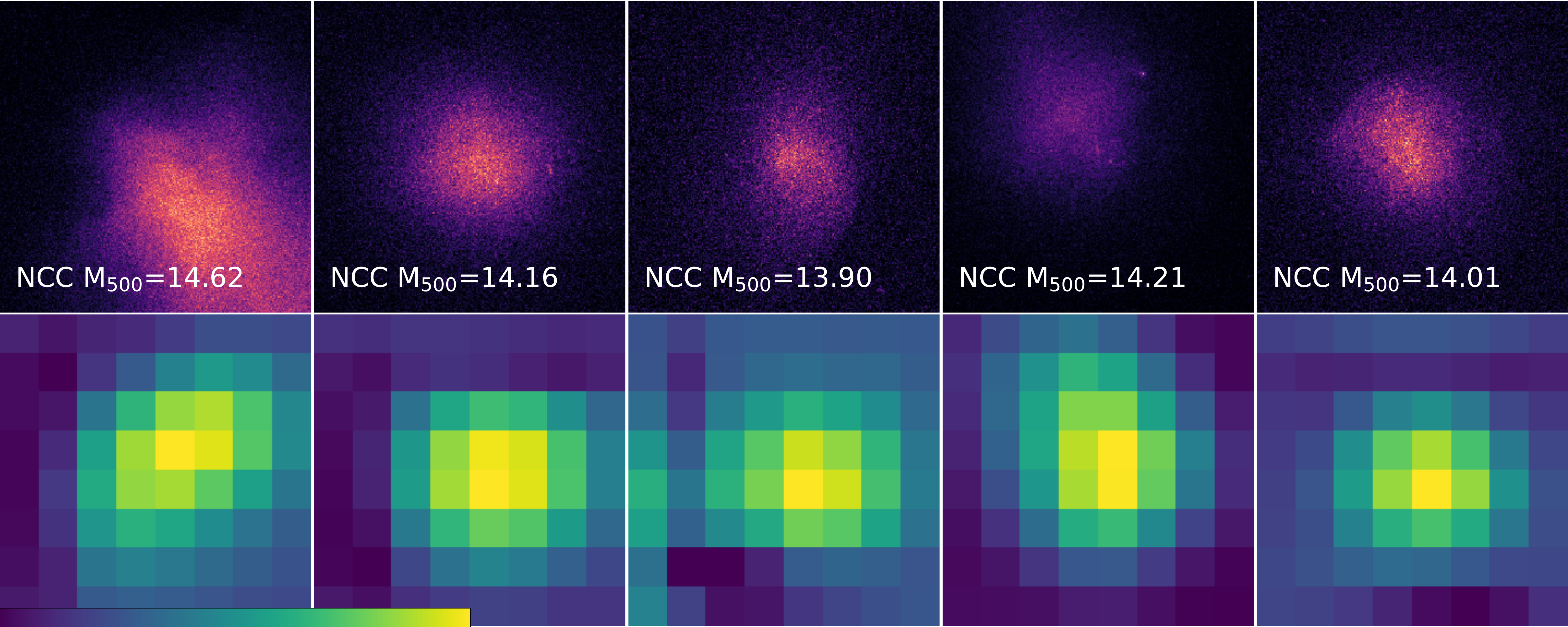}	
   \caption{Same as Figure~\ref{fig:cc_cam} but for non cool core clusters. The network has utilized regions out to the edge of the input image to identify NCC clusters.}
    \label{fig:ncc_cam}
\end{figure*}

\subsection{Mock Chandra observations}
\label{sec:Chandra}

Mock {\sl Chandra} X-ray observations of the TNG300 clusters
are produced in an end-to-end fashion using pyXSIM v2.2.0\symfootnote{\url{http://hea-www.cfa.harvard.edu/~jzuhone/pyxsim}}, an implementation of the PHOX algorithm \citep{ZuHone14, Biffi13}, 
and the
SOXS v2.2.0\symfootnote{\url{http://hea-www.cfa.harvard.edu/soxs}} software suite for simulating X-ray events and producing mock observations. 
A large number of photons in the energy band of 0.5--7.0 keV are generated with pyXSIM for each cluster over a spherical volume with a radius of 2\,Mpc, based on their 3D distributions of density, temperature, and metallicity in TNG300. 
We adopt a ${\tt wabs}\times{\tt apec}$ model, where the ${\tt apec}$ thermal emission model \citep{Foster12} represents the ICM component and the ${\tt wabs}$ model \citep{Morrison83} characterizes the foreground Galactic absorption assuming a hydrogen column density of $4\times10^{20}$\,cm$^{-2}$. 
We assume all the clusters reside at a redshift of $z = 0.05$, such that 1\arcsec=1.01\,kpc for the assumed cosmological parameters in IllustrisTNG. 
Each dataset is then projected along three orthogonal directions x, y, z. 
Mock {\sl Chandra} ACIS-I event files are produced by convolving each photon list with an instrument model for the ACIS-I detector of {\sl Chandra}.
The effective area and spectral response are based on
the Cycle 0 response files. 
The ACIS-I particle background, the galactic foreground, and the Cosmic X-ray background are also included.
Each mock observation is integrated for an exposure time of $100$\,ksec.
We extract images of the central 16.8\arcmin square region in the 0.5--7.0 keV energy band from the simulated event files. The field of view corresponds to a 1 Mpc square at the assumed redshift.     
Each $8\times8$ pixel square is binned up into a single pixel such that the final mock ACIS-I images have a dimension of $256\times256$.
Example mock {\sl Chandra} images of CC, WCC, and NCC clusters are shown in Figures~\ref{fig:cc_cam}--\ref{fig:ncc_cam}. 

\subsection{Neural Network Architecture}
\label{sec:CNN} 

Convolutional Neural Networks \citep[CNNs, ][]{fukushima1982neocognitron, lecun1999object, NIPS2012_4824, simonyan2014deep} are a class of deep machine learning algorithms that are commonly used for image analysis.  
Unlike traditional (shallow) image understanding methods, CNNs extract meaningful patterns from the input imagery using sets of convolutional layers (Conv-layer) with weights that are optimized for a given loss function.
The output of each convolutional layer is a feature map, which is a vector-valued spatial function defined over a grid of image locations. 
Network architectures typically
consist of a linear sequence of Conv-layers followed by a set
of fully connected layers. The Conv-layers extract spatial
features, often with a reduction in spatial resolution later in the
sequence. After the Conv-layers, the spatial feature map is
converted into vector, by either averaging the features across the
image or just reshaping the feature map into a vector
("flattening").
The subsequent fully connected layers label the data with discrete labels for classification tasks or continuous labels for regression tasks. Increasing the number of layers in a CNN will tend to improve results, but at some point, very deep models become too difficult to train. Residual neural networks~\citep[ResNets][]{he2016deep, 2016arXiv160305027H} are a type of CNNs that 
use skip connections, which has been shown to reduce the difficulty in training CNNs with many layers.
ResNets have been used in astronomical applications including finding strong gravitational lenses~\citep{2018MNRAS.473.3895L}, galaxy morphology classification~\citep{2019Ap&SS.364...55Z}, and identifying candidate Lyman-$\alpha$ emitting galaxies~\citep{2019MNRAS.482..313L}.

\begin{figure}
    \centering
    \includegraphics[width=0.4\textwidth]{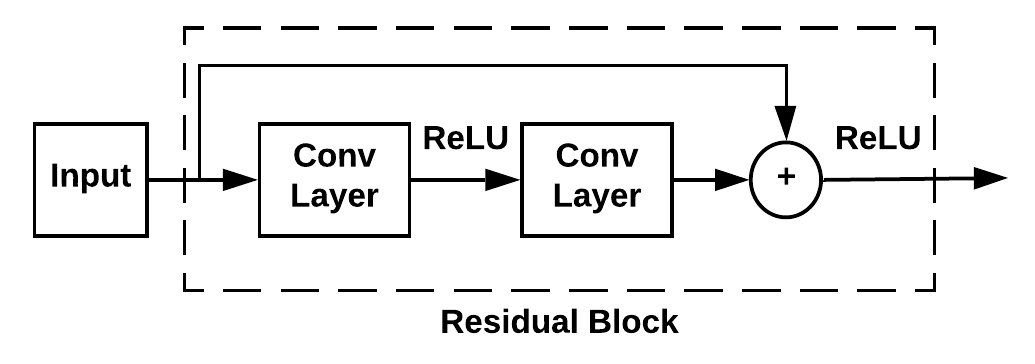}
    \caption{A residual block is a two Conv-layers shallow network with each Conv-layer followed by a ReLU. A skip connection passes the input of the residual block to be added to the output of the second Conv-layer and fed to the second ReLU. One advantage of building a model with residual blocks is that it allows for deeper and more flexible models that can be trained efficiently.}
    \label{fig:res_block}
\end{figure}

\begin{figure*}
    \centering
    \includegraphics[width=0.95\textwidth]{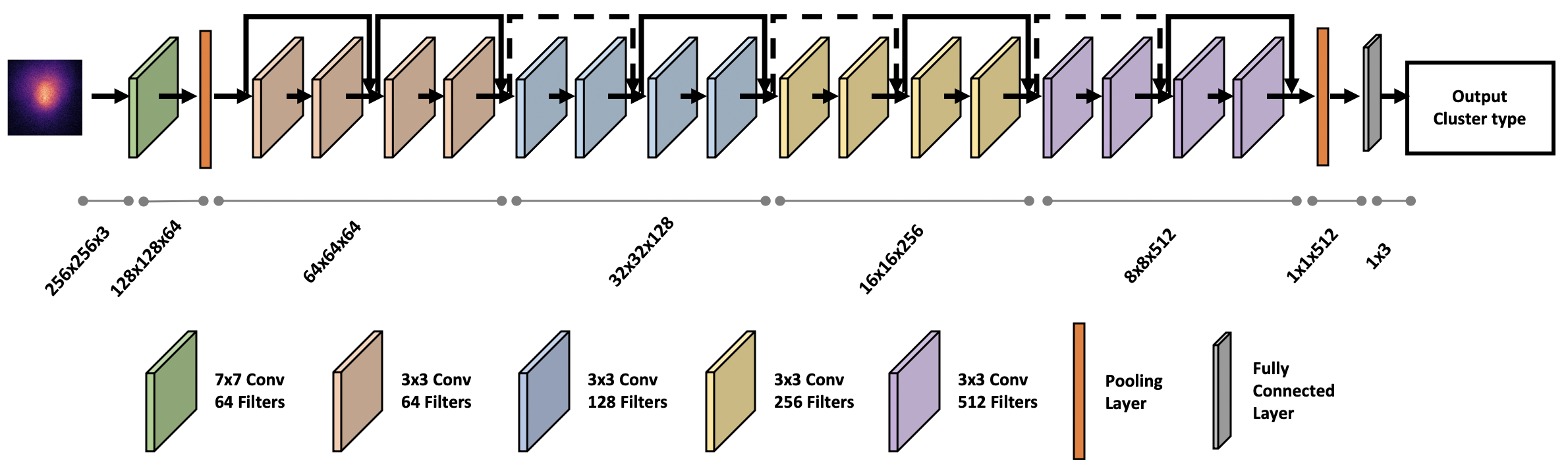}
    \caption{Architecture of a ResNet-18 neural network. The input and output shapes of each layer are labeled. After the first pooling layer, every two Conv-layers form a residual block as illustrated in Figure~\ref{fig:res_block}. The dashed shortcuts involve dimension changes.}
    \label{fig:architecture}
\end{figure*}

The ResNet-18 network, employed in this study, contains one Conv-layer, 8 residual blocks, and one fully connected layer. 
A residual block is a shallow network of two Conv-layers (Figure~\ref{fig:res_block}). Each Conv-layer is followed by a Rectified linear unit (ReLU)~\citep{zeiler2013rectified}. A skip connection is added to the data passing flow to directly pass the input of the residual block to the end of the second Conv-layer. The input of the residual block and the output of the second Conv-layer are then added together to be fed to the second ReLU. The output of the second ReLU is the output of the residual block. 
A $3\times3$ max-pooling layer follows the first Conv-layer, and a global average pooling (GAP) layer follows the last residual block. 
The ResNet-18 network contains a total of 18 hidden layers. Its basic architecture is shown in Figure~\ref{fig:architecture}. 

Our network is implemented in PyTorch~\citep{paszke2019pytorch}. A learning rate of ${\rm lr}=0.001$, a batch size of $64$, and Adam optimizer~\citep{kingma2014adam} are used during training. 
A ResNet-18 model is pre-trained on the ImageNet Dataset which contains over one million images
for a 1000-class classification~\citep{deng2009imagenet}. The pre-trained network is fine tuned with our dataset for predicting cluster types.
Weighted cross-entropy~\citep{lecun2015deep} is used as our loss function. Weights that are inversely proportional to the number of data in each class are included in the loss function to mitigate the impacts of the imbalanced dataset.
The mock {\sl Chandra} images have a dimension of $256\times256$. 
Since the ResNet-18 network expects a 3-channel input image, each image is replicated three times to form a $256\times256\times3$ image. 
All the input images are randomly split into 10 folds (groups) of roughly equal size. No image from the same cluster appears in more than one fold. 
We use 8 folds for training, 1 fold for validation, and 1 for testing. Input images are augmented by a random combination of horizontal/vertical flip and 0/90/180/270 degrees rotation during training. 
Each model is trained for 200 epochs. 
The model that gives the highest F$_1$-score (Equation~\ref{eq:f1}) on the validation set is chosen and used for testing. 
A 10-fold cross-validation has been applied to cycle through all the data. 

We apply a Class Activation Mapping \citep{inproceedings} technique to highlight regions that are discriminative for the CNN. We compute a weighted sum of the feature maps of the last Conv-layer to obtain a class activation map (CAM) for each image. 
The activation of unit $k$ in the last Conv-layer at a 2D coordinate of $(x,y)$ is $f_k(x,y)$. The result of global average pooling for that unit is $F_{k}=\sum_{x,y}f_k(x,y)$. The input to the softmax for class $c$ is $S_c=\sum_{k}w^c_kF_k$, where $w^c_k$ is the weight for class $c$ and unit $k$. We obtain 
\begin{equation}
S_c=\sum_{x,y}\sum_{k}w^c_kf_k(x,y)
      =\sum_{x,y}M_c(x,y)
\end{equation}
where $M_c(x,y)$ is the value on the CAM for position $(x,y)$.
The probability for each class, $P_c={\rm exp}(S_c)/\sum_c{\rm exp}(S_c)$, is used to make the final decision. CAM therefore reveals the importance of each part in an image that leads to the classification of an image to a class. 
The resulting CAM has the same dimension as the output of the last Conv-layer and the input of the GAP layer of 8x8.

\section{Results}
\label{sec:results}
We use our CNN algorithm to predict whether a cluster is CC, WCC, or NCC from the mock {\sl Chandra} X-ray images. The cluster types are defined by their actual central cooling times in TNG300. 
We compare the performances with the estimates given by two traditional methods of using central gas densities and surface brightness concentrations.  

We use the following criteria to evaluate the performance of each experiment. Hereafter, $tp$, $fp$, $tn$, and $fn$ are the numbers of true positive, false positive, true negative, and false negative predictions, respectively. 
Precision, also called positive predictive value, is the number of true positives, divided by the number of all positive calls:
\begin{equation}
{\rm Precision}=\frac{tp}{tp+fp}.
	\label{eq:precision}
\end{equation}
Recall, also called true positive rate, is the number of true positives divided by the number of positive samples:
\begin{equation}
{\rm Recall}=\frac{tp}{tp+fn}.
\label{eq:recall}
\end{equation}
F$_1$-score is the harmonic mean of precision and recall, 
defined as:
\begin{equation}
{\rm F_1}=2 \cdot\frac{\rm Precision \cdot Recall}{\rm Precision+Recall}.
\label{eq:f1}
\end{equation}
{It conveys the balance between precision and recall and provides a more comprehensive evaluation. We base our main conclusions on F$_1$-score.} 
Balanced accuracy (BAcc) is the average of true positive predictions divided by the number of positive samples and true negative predictions divided by the number of negative samples.
It is related to $tp$, $fp$, $tn$, and $fn$:
\begin{equation}
{\rm BAcc}=\frac{1}{2}(\frac{tp}{tp+fn}+\frac{tn}{tn+fp}). 
	\label{eq:accuracy}
\end{equation}
BAcc is a measurement of accuracy that does not suffer from imbalanced datasets.

\begin{figure*}
	\includegraphics[width=0.45\textwidth]{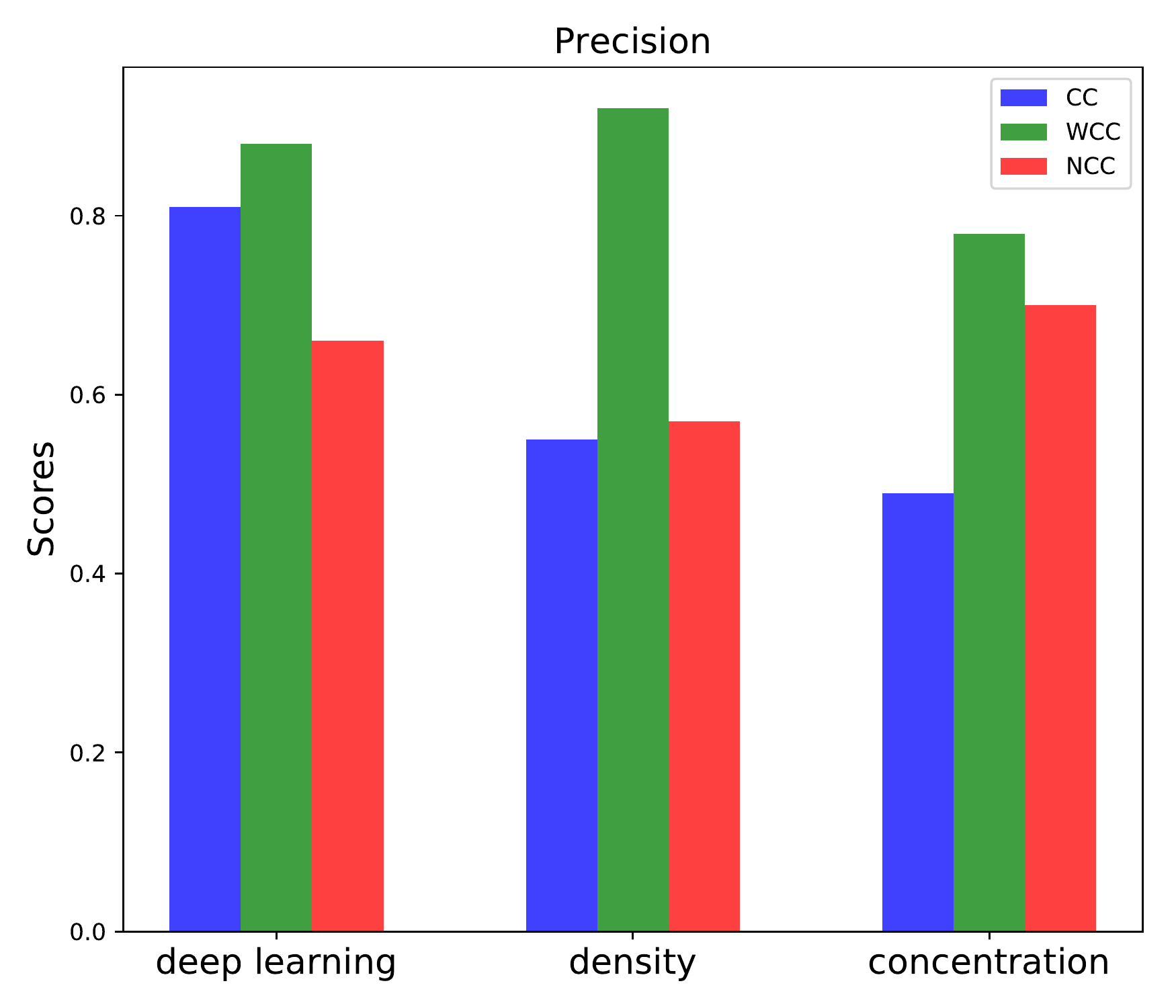}
	\includegraphics[width=0.45\textwidth]{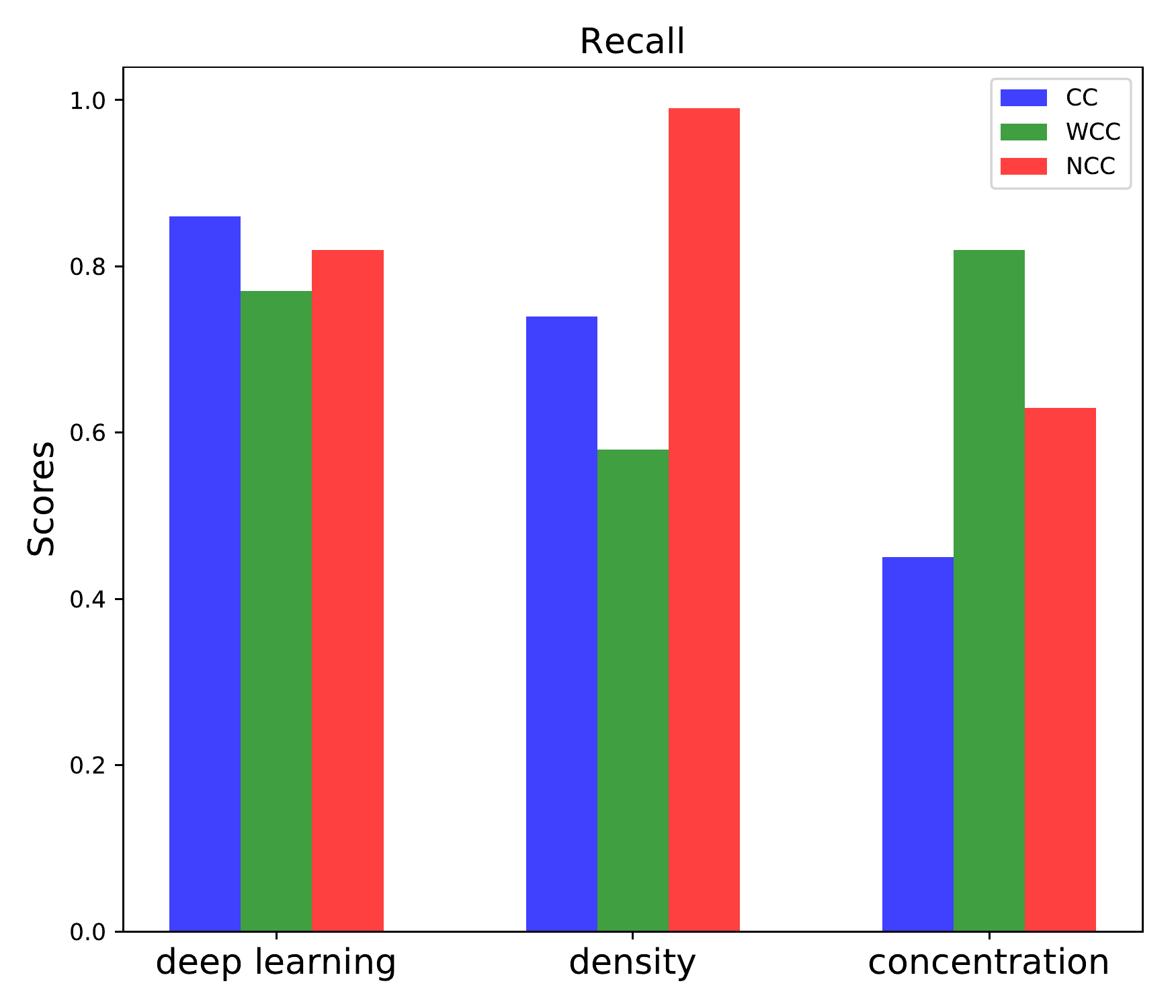}
	\includegraphics[width=0.45\textwidth]{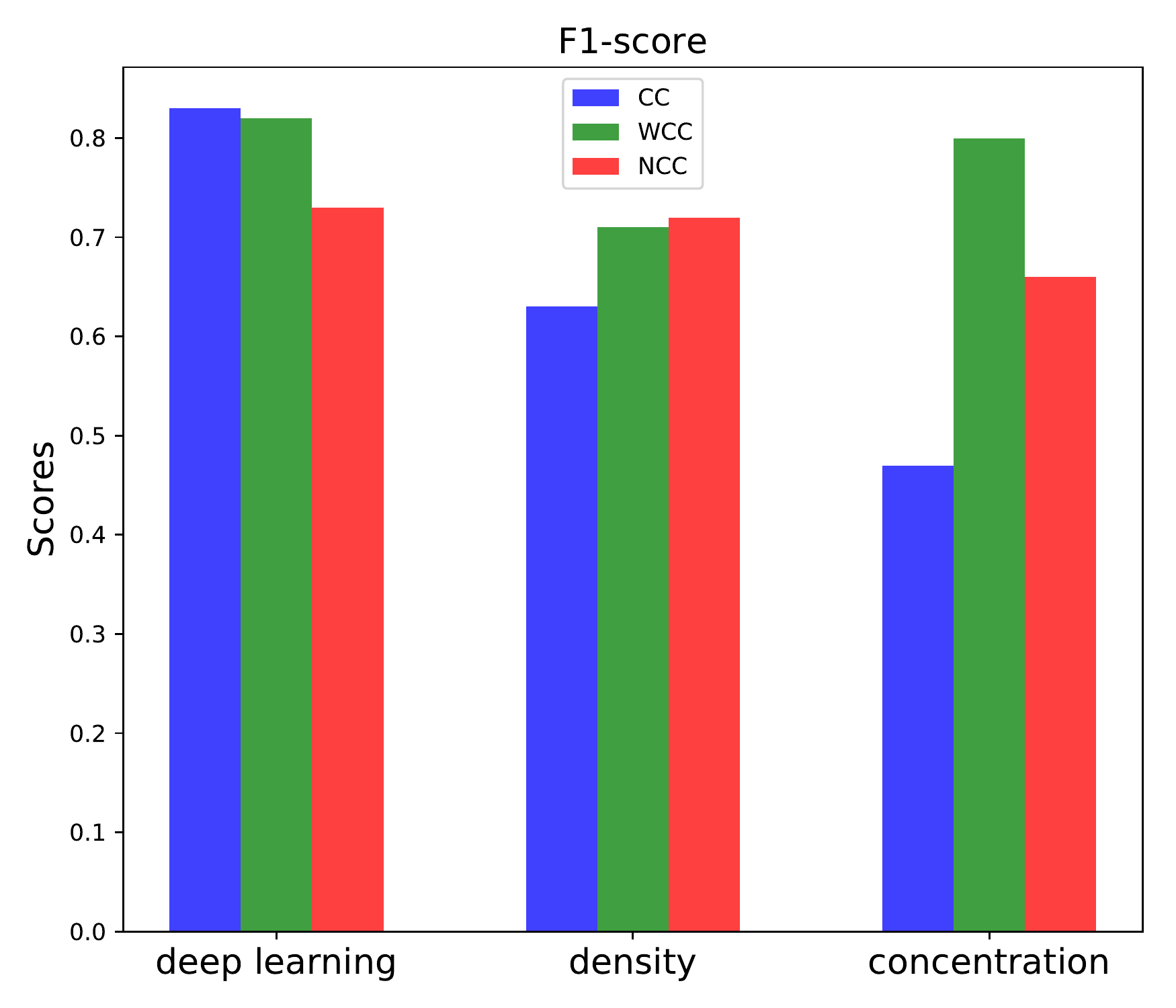}
	\includegraphics[width=0.45\textwidth]{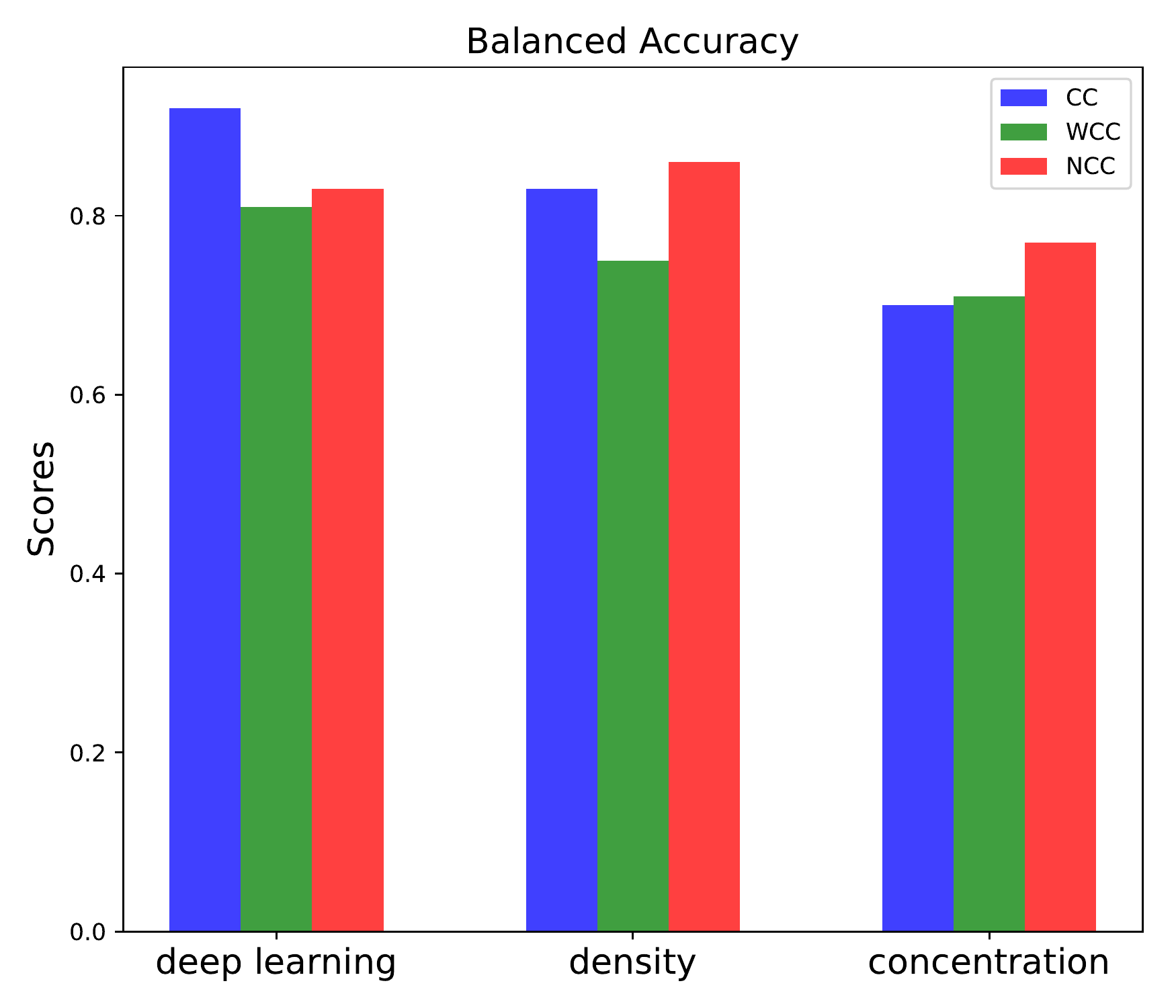}
    \caption{Scores of precision (top-left), recall (top-right), F$_1$-score (bottom-left), and balanced accuracy (bottom-right) for CC, WCC, and NCC classifications obtained using different methods (x-axis): 1. deep learning using mock {\sl Chandra} images, 2. central gas density, 3. X-ray concentration parameter derived from mock {\sl Chandra} images. Details of the performance measures are listed in Table~\ref{tab:table}.}
    \label{fig:bar}
\end{figure*}

\begin{figure*}
	\includegraphics[width=1.0\textwidth]{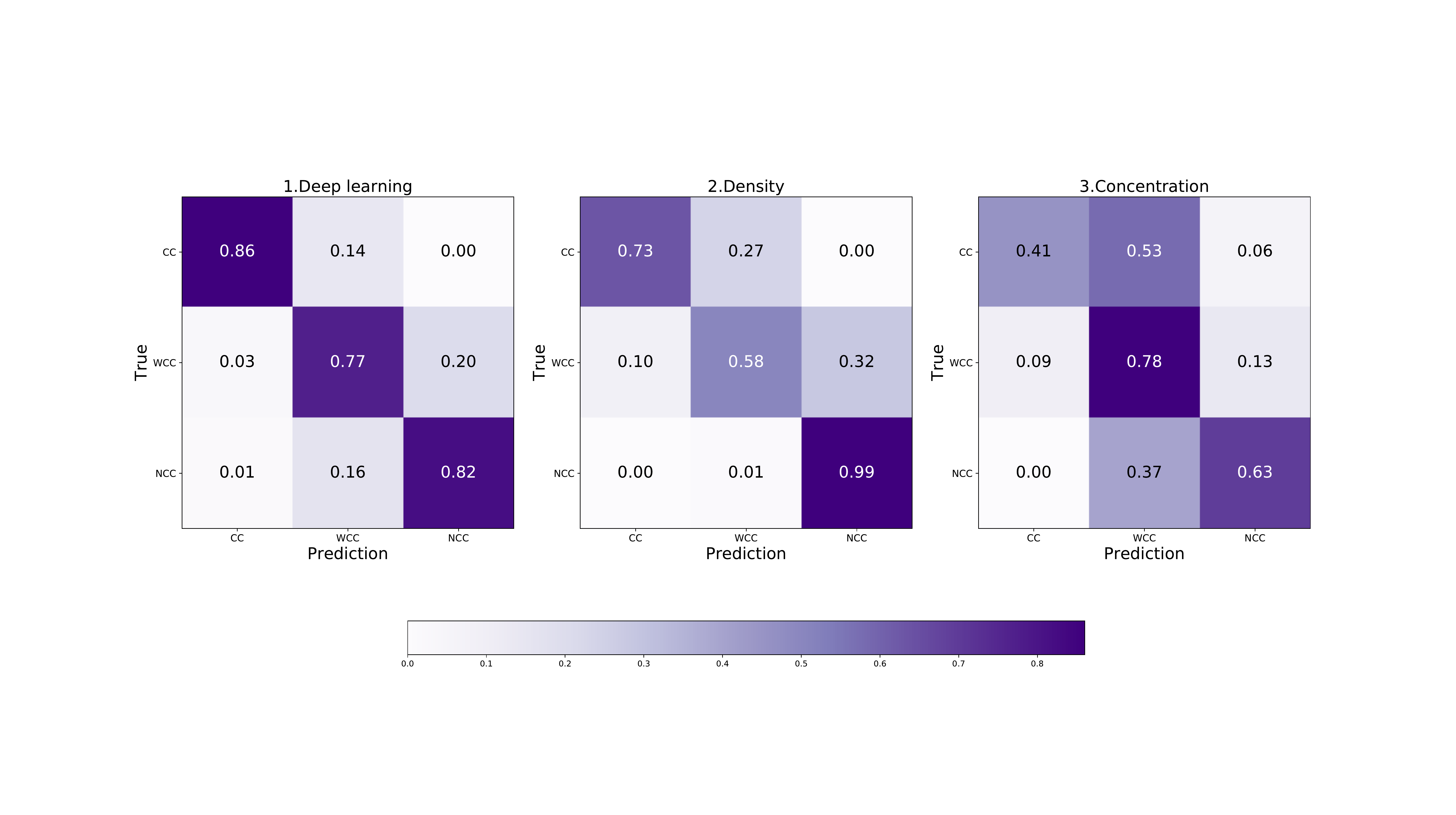}
    \caption{Normalized confusion matrix of CC, WCC, and NCC classification which compares
the predicted class and the true class for each experiment. 1. deep learning using mock {\sl Chandra} images, 2. central gas density, 3. X-ray concentration parameter derived from mock {\sl Chandra} images. Details of the performance measures are listed in Table~\ref{tab:table}.}
    \label{fig:matrix}
\end{figure*}

We train and test our deep learning classification algorithm with mock {\sl Chandra} ACIS-I images as shown in Figures~\ref{fig:cc_cam}--\ref{fig:ncc_cam} and described in \S\ref{sec:Chandra}. 
Each ACIS-I field covers a 1\,Mpc square,
whereas clusters in our sample have a median $R_{500}$ of 710\,kpc.
The spatial resolution is degraded to 3.9\arcsec/pixel which is 8 times worse than the half arcsec resolution of {\sl Chandra} ACIS. 
Without any spectral information, the network is able to distinguish 
CC, WCC, and NCC clusters with F$_1$-scores of 0.83, 0.82, and 0.73, respectively. 
Details of the results are shown in Figure~\ref{fig:bar} and values of performance measures are listed in Table~\ref{tab:table}. Predictions and the ground truths are compared in the normalized confusion matrix as shown in Figure~\ref{fig:matrix}. 
Diagonal elements represent the fraction of data for which the predicted class is the same as the true class, while off-diagonal elements are those that are misclassified. The deep learning algorithm gives a confusion matrix with high diagonal values, indicating good predictions.  

{Here we compare the deep learning method to more conventional methods for cluster classification. Although these approaches are not directly comparable, the comparisons are instructive.}  
A rapidly cooling core implies a high central gas density as the ICM gradually loses its pressure support and falls to smaller radii. 
Central gas densities have been widely used to determine whether a cluster contains a cool core, which requires far fewer counts than measuring the temperatures and metallicities \citep{lorenzo15,Su19}. 
We calculate the central electron number density $n_e$ as the 
average $n_e$ of a 3D volume within $0.012\,R_{500}$ as shown in \citet{Barnes18}.
Following \citet{Barnes18} and \citet{Hudson10}, clusters with a central $n_{\rm e}>1.5\times10^{-2}$\,cm$^{-3}$, $1.5\times10^{-2}>n_{\rm e}>0.5\times10^{-2}$\,cm$^{-3}$, and $n_{\rm e}<0.5\times10^{-2}$\,cm$^{-3}$ are classified as CC, WCC, and NCC, respectively. 
For clusters in our sample, this method achieves ${\rm F}_1=0.69$, averaged over the three cluster types (see Figures~\ref{fig:bar} and \ref{fig:matrix} and Table~\ref{tab:table}),
which is not as accurate as the predictions given by our ResNet-18 classifier.

In X-ray observations, it is challenging to directly measure gas properties within $r\lesssim10$\,kpc for a modest exposure time.
The elevated ICM metallicity and density at the centre of a CC cluster produce a central peak in X-ray surface brightness. 
The ratio of this peak emission to the ambient emission is therefore sensitive to the cool core strength. 
The X-ray concentration parameter was originally introduced by \cite{2008A&A...483...35S} to infer whether a cluster contains a cool core:
\begin{equation}
C_{\rm SB}=\frac{\sum(<40{\rm kpc})}{\sum(<400{\rm kpc})}
	\label{eq:csb}
\end{equation}
where $\sum(<r)$ is the accumulated projected ICM emission in 0.5--5.0 keV from a circular region 
with a radius of $r$. 
We extract images in the 0.5--5.0 keV energy band from mock {\sl Chandra} observations. 
Following \citet{Barnes18} and \citet{Felipe17}, 
clusters with $C_{\rm SB}> 0.155$,  $0.075<C_{\rm SB}< 0.155$, and $C_{\rm SB}< 0.075$ are classified as CC, WCC, and NCC, respectively.
Using this method, we obtain an average F$_1$-score of 0.33 for clusters in our sample.
\citet{Barnes18} also note that this criterion overpredicts NCC clusters and fails to identify CC clusters. 
We further sort all the images with a decreasing $C_{\rm SB}$ and 
divide them into the three categories based on the fractions of CC, WCC, and NCC in our sample.
We obtain an F$_1$-score of 0.64 (see Figures~\ref{fig:bar} and \ref{fig:matrix} and Table~\ref{tab:table}).  
Our ResNet-18 classifier which utilizes the 2D ICM distribution outperforms the 1D concentration measurement.

\begin{table*}
	 \captionsetup{justification=centering}
	 \centering
	\caption{Values of performance measures for different experiments. Precision, recall, F$_1$-score, and balanced accuracy are defined in \S\ref{sec:results}.}
\begin{tabular}{cccccccc}
 \hline
method & ave. F$_1$ & ave. BAcc  & class & precision & recall & F$_1$ & BAcc  \\
 \hline
 &  &  & CC & 0.81 & 0.86 & 0.83 & 0.92 \\
deep learning & 0.79 & 0.85 & WCC & 0.88 & 0.77 & 0.82 & 0.81 \\
 &  &  & NCC & 0.66 & 0.82 & 0.73 & 0.83 \\
  \hline
 &  &  & CC & 0.55 & 0.74 & 0.63 & 0.83 \\
density & 0.69 & 0.81 & WCC & 0.92 & 0.58 & 0.71 & 0.75 \\
 &  &  & NCC & 0.57 & 0.99 & 0.72 & 0.86 \\
  \hline
 &  &  & CC & 0.49 & 0.45 & 0.47 & 0.70 \\
concentration & 0.64 & 0.73 & WCC & 0.78 & 0.82 & 0.8 & 0.71 \\
 &  &  & NCC & 0.70 & 0.63 & 0.66 & 0.77 \\
  \hline
\end{tabular}
 \label{tab:table}
\end{table*}

\begin{figure*}
	\includegraphics[width=0.69\textwidth]{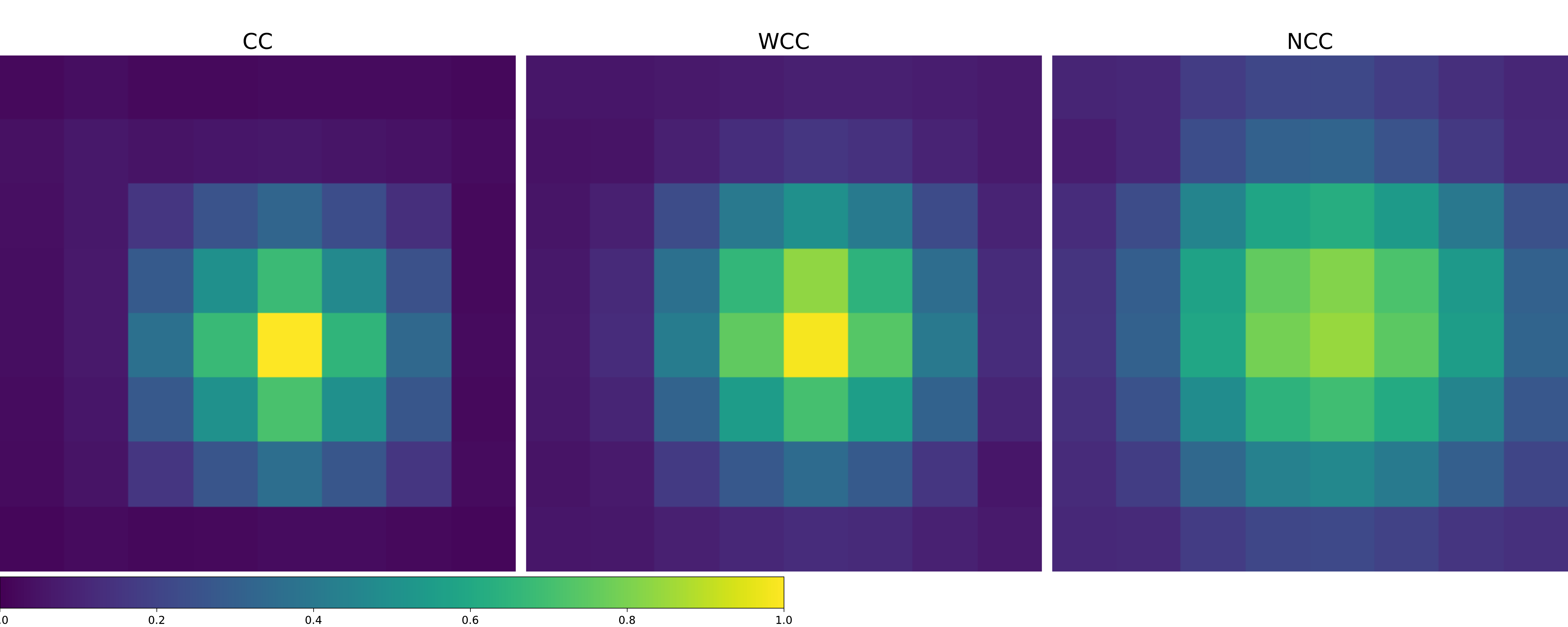}
	\includegraphics[width=0.28\textwidth]{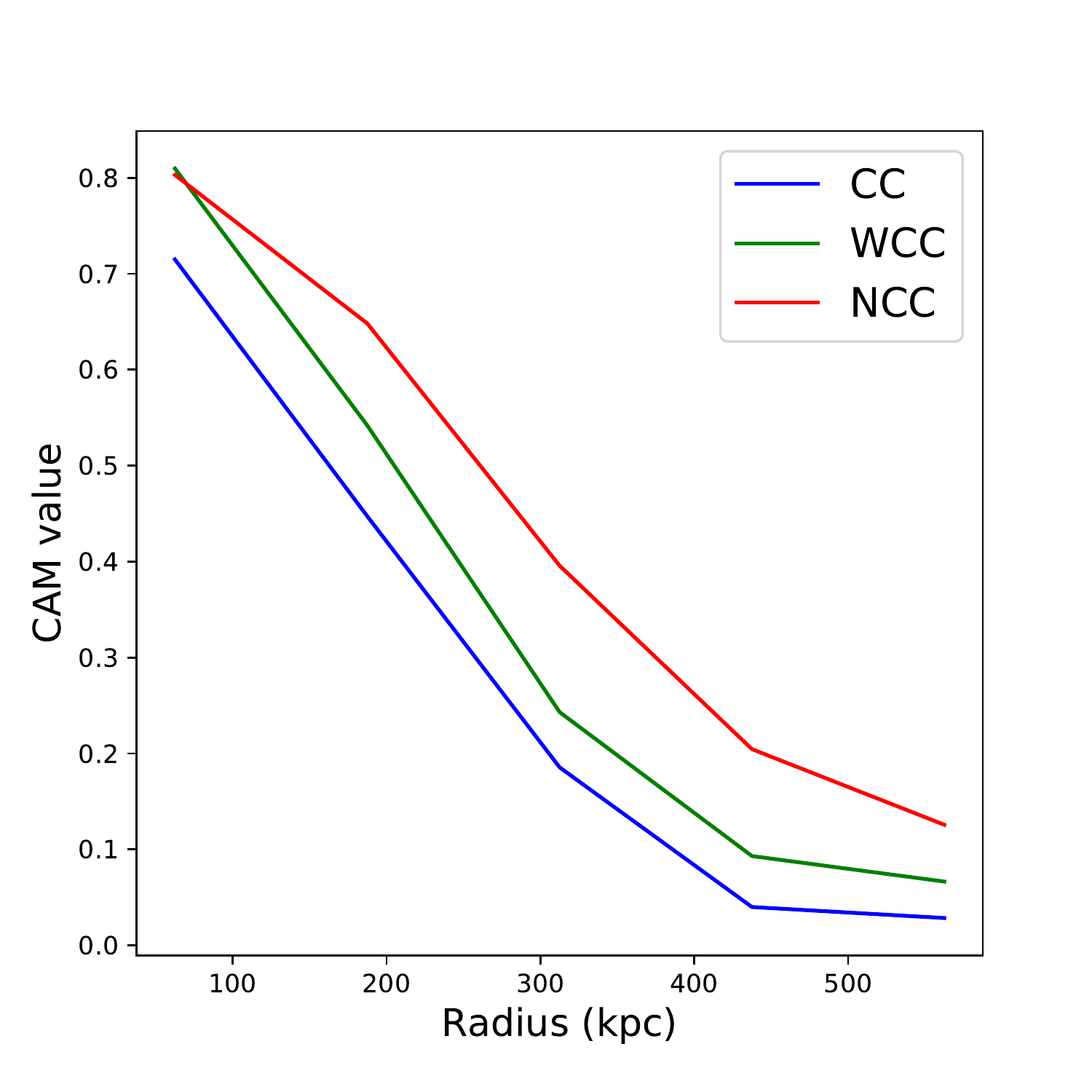}		
    \caption{Class activation maps of the central $D=1$\,Mpc averaged over cool core, weak cool core, and non cool core clusters. respectively. All these clusters are predicted correctly with a probability above 0.9. 
The right panel shows the radial profiles of the three CAM maps.     
The network utilizes relatively more information from the cluster centres to identify CC clusters but relies on the morphology over a wider radial range to identify NCC clusters. The radial dependance of the discriminating power of regions in WCC clusters is between those of CC and NCC clusters.}
    \label{fig:cam}
\end{figure*}

\section{Interpretation and discussion}
\label{sec:discussion}

Using mock {\sl Chandra} X-ray images of a 1\,Mpc square centred on each cluster, our network is able to predict whether a cluster is CC, WCC, or NCC with an average F$_1$-score of 0.79. 
The cluster types are defined by their actual central cooling times, which depend on temperature, density, and metallicity as shown in Equation~\ref{eq:tc}. Our deep learning method is superior to the estimate using the actual central gas densities of these clusters with an average ${\rm F_1}=0.69$. X-ray images may contain information that is more directly related to the cooling time than gas density. 
To localize features that are most useful for the network to make classification decisions, we generate a class activation map (CAM) for each input image as described in \S\ref{sec:CNN}. Example CAM images and their original images are compared in Figures~\ref{fig:cc_cam}--\ref{fig:ncc_cam}. Regions that are brighter in CAM are more informative for the network.
We stack and normalize all the CAM images associated with correct predictions with a probability above 0.9 for CC, WCC, and NCC clusters, respectively, as shown in Figure~\ref{fig:cam}.
The radial profiles of the values in the activation maps are shown in the
right panel of Figure~\ref{fig:cam}. 
To identify CC clusters, the network uses 2D information out to $r\approx300$\,kpc which is broader than $r<0.012\,R_{500}$ (10\,kpc) where the central gas density and cooling time are measured.
Patterns in X-ray images could provide important clues about the central cooling time. For example, sizes of X-ray cavities and their distances to the cluster centre are determined by the outburst of the AGN, which is related to the radiative cooling rate \citep{Birzan12,  2018MNRAS.480.4279L}. 
Mechanical energies released by AGN could be dissipated by heating the ICM via turbulent cascades, which can be probed through the power spectrum of X-ray surface brightness fluctuations \citep{Zhuravleva14}.   
These informative 2D features in X-ray images may have allowed the network to 
obtain stronger constraints on $t_{\rm cool}$ than the methods of using central gas densities and surface brightness concentrations.

While the discriminating power of each region declines quickly as a function of radius for CC clusters, it is relatively uniform for NCC cluster. 
The network uses extended regions out to the edge of the input images of $r\sim500$\,kpc to identify NCC clusters as shown in Figure~\ref{fig:cam}. Mergers may have disrupted the thermal structures of the ICM and contributed to the formation of NCC clusters. 
Interestingly, \citet{Barnes18} find that NCC clusters do not have significantly higher kinetic energies than CC clusters. We speculate that head-on mergers and certain off-axis mergers may have similar impacts on the global kinetics of the clusters.
Cluster cool cores are resilient to off-axis mergers as indicated by the ubiquitous presence of sloshing cold fronts in CC clusters~\citep{2007PhR...443....1M, 2017ApJ...851...69S}.
The impact parameter and angular momentum of a merger may be critical in determining the fate of cluster cool cores, as seen in numerical simulations~\citep{2017MNRAS.470..166H}.   

{We note that the misclassified cases are either associated with cooling times that are close to the class boundaries or outliers for their class. Some of the failures in our model are shown and discussed in detail in Appendix~\ref{sec:mis}.
We only considered clusters at a single redshift in this work. Our algorithm may have some resilience to distance since
the dataset consists of clusters with a wide range of masses and physical sizes. The application of deep learning techniques to systems at different redshifts will be an important aspect of future studies.}  
In overall, this work demonstrates that CNNs are able to take advantage of X-ray images and provide a unique approach to the thermodynamics of the ICM.  
The neural network can, in principle, be trained to predict the specific values of $t_{\rm cool}$ for CC clusters as a regression task, and fetch features associated with the life cycles of AGN feedback.
Potentially, the deep learning algorithm can also be used to determine the merger history of a cluster from multi-wavelength images -- X-ray, radio, and optical, which would greatly enhance our understanding of cluster formation, thermalization, and particle acceleration. 

\section{Conclusions}
\label{sec:conclusion}
ResNet-18 is a subclass of convolutional neural networks that is well suited for image classification. 
We employ a ResNet-18 network to assess whether a cluster is CC, WCC, or NCC from their X-ray images. The cluster type is defined purely by its central cooling time, which is related to the gas density, temperature, and metallicity. 
{We produce mock {\sl Chandra} observations for 318 clusters of galaxies in TNG300 with particle background, contaminating point sources, galactic foreground, etc. included.}
We train and test the network with low resolution mock {\sl Chandra} ACIS-I images.  
It achieves an average precision, recall, F$_1$-score, and balanced accuracy of 0.78, 0.82, 0.79, and 0.85, respectively, well above a random prediction of 0.33. 
Our deep learning algorithm outperforms the estimates given by the central gas densities and surface brightness concentration parameters. 
We use the class activation mapping to probe the contribution of each region to the classification decisions. 
The network may have utilized 2D features in X-ray images that are related to the cooling and heating mechanisms in the intracluster medium. Features at larger radii are more important for identifying NCC clusters than CC clusters, possibly due to the role of head-on major mergers in disrupting cluster cool cores.

Unlike traditional methods of using one dimensional information to estimate the cluster type, the neural network is able to identify features on different scales and at various radii, making it a potentially powerful tool to probe the thermodynamic state of a cluster. 
CNNs can be utilized to exploit cluster images in {\sl Chandra} and {\sl XMM-Newton} archives, large cluster samples from the ongoing eROSITA all sky survey, and the exquisite data promised by next-generation X-ray observatories, such as {\sl Lynx}.

\section*{Acknowledgements}
The authors thank the anonymous referee for his/her helpful comments.
We acknowledge the use of the Lipscomb High-Performance Computing Cluster at the University of Kentucky for conducting this research. 
Y.\ S. was partially supported by Chandra X-ray
Observatory grants AR8-19020A and GO6-17125A. 
\section*{Data availability}
The data underlying this article will be shared on reasonable request to the corresponding author.



\bibliographystyle{mnras}
\bibliography{main_accepted.bib} 






\appendix

\section{Misclassified clusters}
\label{sec:mis}
\begin{figure*}
 \centering
	\includegraphics[width=0.6\textwidth]{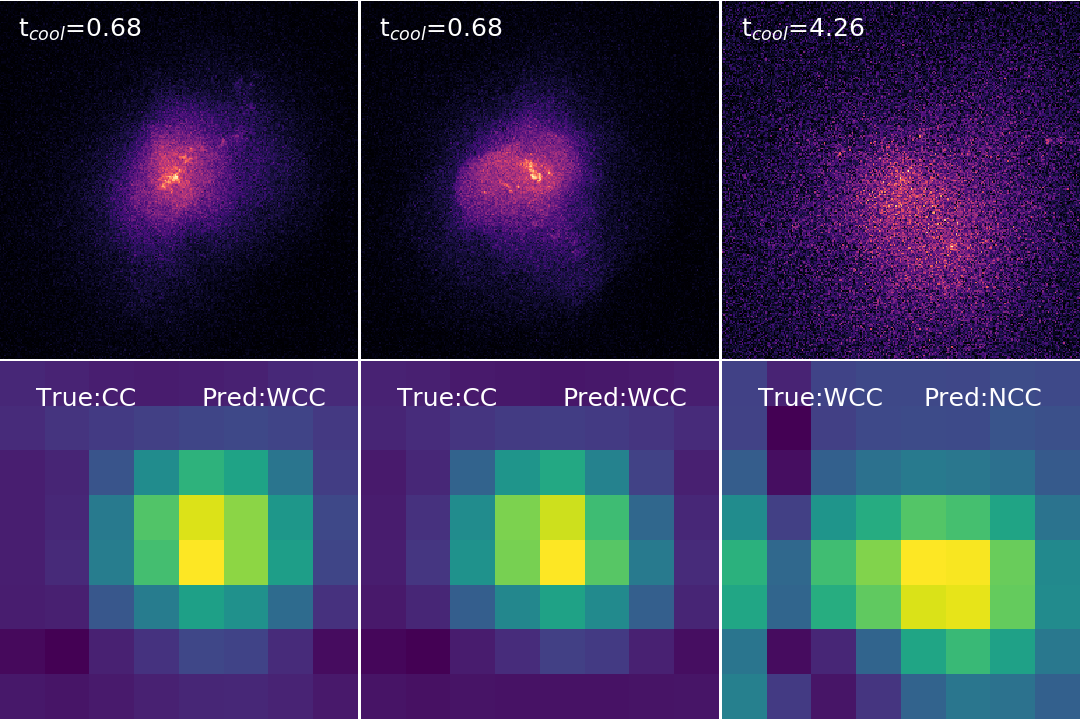}
	
	\vspace{0.35 cm}

	\includegraphics[width=0.6\textwidth]{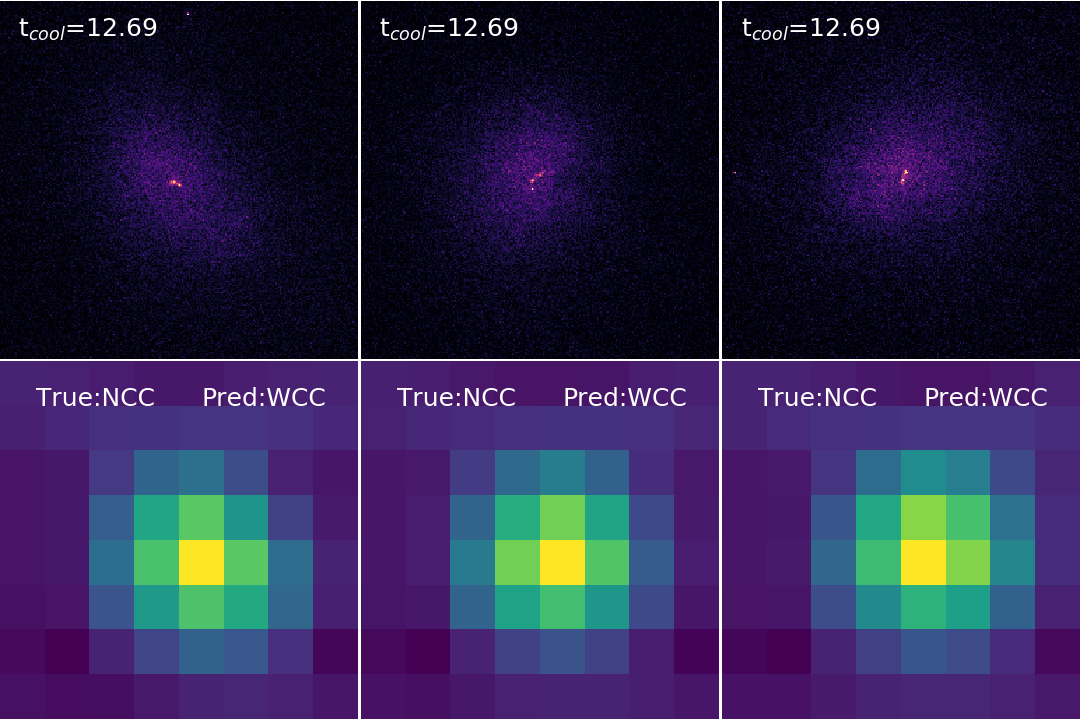}		
    \caption{Same as Figures~\ref{fig:cc_cam}--\ref{fig:ncc_cam} but for clusters that are misidentified by the neural network.
    Their true cooling times are labeled in the {\sl Chandra} X-ray images and their true and predicted cluster types are labeled in the class activation maps. All these clusters are predicted incorrectly with a probability above 0.9. Their X-ray morphologies appear to be atypical for their cluster types.}
    \label{fig:mis}
\end{figure*}

{To better understand the failures of our model, we inspect the cases that are classified confidently but incorrectly.
All the incorrect predictions with a probability above 0.9 are associated with WCC, which supports that WCC is a transitional phase between CC and NCC with intermediate morphologies that are more difficult to classify.  
Among them, 
2 images are CC classified as WCC, 8 are NCC classified as WCC, 4 are WCC classified as CC, and 8 are WCC classified as NCC. Most of these clusters have cooling times close to the boundaries of $t_{\rm cool} \left(\rm CC|WCC\right)=1$\,Gyr and $t_{\rm cool} \left(\rm WCC|NCC\right)=7.7$\,Gyr. However, 
6 images from 3 clusters have typical coolings times for their type, as shown in Figure~\ref{fig:mis}. These cases appear to be outliers in their class. 
The top-left and top-middle images in Figure~\ref{fig:mis} are from the same CC but it appears to be disturbed with a very asymmetric morphology and the network classified it as WCC. The top-right image is a WCC and it appears to be undergoing a merger. 
CAM suggests that the network may have noticed the subcluster in the lower right corner and classified it as NCC.
The three images in Figure~\ref{fig:mis}-bottom are from the same NCC cluster but classified as WCC. It appears to be relatively relaxed. The double nuclei at the cluster center may be mistaken as a bright filament.  
The network may not be well trained to identify these outliers due to the rarity of such atypical cases in our sample.}

\bsp	
\label{lastpage}
\end{document}